\documentclass[aps,prb,twocolumn,superscriptaddress,showpacs]{revtex4}
\usepackage{bm}
\usepackage{graphicx}

\begin{document}


\title{Experimental manifestation of the breakpoint region in the current-voltage characteristics of intrinsic Josephson junctions}


\author{A. Irie}
\affiliation{Department of Electrical and Electronic Systems Engineering, Utsunomiya University, 7-1-2 Yoto, Utsunomiya 321-8585, Japan}
\author{Yu. M. Shukrinov}
\affiliation{Bogoliubov Laboratory of Theoretical Physics, Joint Institute for Nuclear Research, Dubna, Moscow Region, 141980, Russia}
\author{G. Oya}
\affiliation{Department of Electrical and Electronic Systems Engineering, Utsunomiya University, 7-1-2 Yoto, Utsunomiya 321-8585, Japan}


\date{\today}

\begin{abstract}
The experimental evidence of the breakpoint on the current-voltage characteristics (IVCs) of the stacks of intrinsic Josephson junctions (IJJs) is presented.
The influence of the capacitive coupling on the IVCs of Bi$_2$Sr$_2$CaCu$_2$O$_y$ IJJs has been investigated.
At 4.2 K, clear breakpoint region is observed on the branches in the IVCs.
It is found that the hysteresis observed on the IVC is suppressed due to the coupling compared with that expected from the McCumber parameter.
Measurements agree well with the results obtained by the theoretical model.
\end{abstract}




\maketitle


It is well known that layered Bi$_2$Sr$_2$CaCu$_2$O$_y$ single crystals represent natural stacks of atomic intrinsic Josephson junctions with interlayer spacing of 1.55 nm. \cite{kleiner:prl1992,oya:jjap1992}
Their physical properties are different from those of conventional Josephson junctions because there exists the inductive or capacitive coupling between adjacent junctions due to extremely thin superconducting CuO$_2$ layer of 0.3 nm.

The influence of the coupling effect on the IVCs has been studied both theoretically and experimentally.
In the absence of an external magnetic filed, the capacitive-coupling may become important because the thickness of CuO$_2$ layers is comparable to the Debye charge screening length $r_D$.
From this point of view, the capasitively coupled Josephson junctions (CCJJ) model was proposed to describe the phase dynamics of IJJs \cite{koyama:prb1996,machida:prl1999} and was used to analyze their IVC.
On the other hand, it was also theoretically shown that the diffusion current plays an important role in the IJJ stack.\cite{ryndyk:prl1998}
Therefore, the CCJJ model including the diffusion current (CCJJ+DC model) may be more suitable for the description of IJJs under no external magnetic field.
Recently, Shukrinov et al. have been systematically studied the phase dynamics of IJJs by using the CCJJ+DC model and predicted that the capacitive-coupling effect induces the appearance of new fine structures, which is called as breakpoint region (BPR), together with an equidistant branch structure in the IVC. \cite{shukrinov:physicaC2006,shukrinov:sust2007,shukrinov:prb2007,shukrinov:prl2007}
The BPR is caused by the parametric resonance between Josephson and plasma oscillations, and is determined as a region between the breakpoint current $I_{bp}$, at which the resonance is happened and the longitudinal plasma wave (LPW) is excited, and the transition current $I_j$ to another branch in the IVC.
However, to our knowledge, the experimental observation of the BPR in the IVC of the IJJs has been not done yet.

In this letter, we show the first experimental results which demonstrate the BPR in the IVC.

Bi$_2$Sr$_2$CaCu$_2$O$_y$ single crystals with critical temperatures $T_c$ of $\sim85$ K were grown by a conventional melting method.
The grown single crystals were glued on glass substrates and the cleaved in air to obtain their fresh surface.
Subsequently, Au thin films of $20-40$ nm thickness were deposited on the cleaved surfaces.
And, the mesas with areas $S$ of $9,~16,~25,~36,~49,~64~\mu$m$^2$, consisting of 5-20 IJJs, were fabricated by using electron beam lithography, photolithography and Ar ion milling.
Their IVC along the $c$-axis direction were measured using a three terminal method.
Here, we present the results for two samples: Nm1-25 and Nm5-25 with $S=25~\mu$m$^2$.

\begin{figure}
\includegraphics[width=0.29\textwidth]{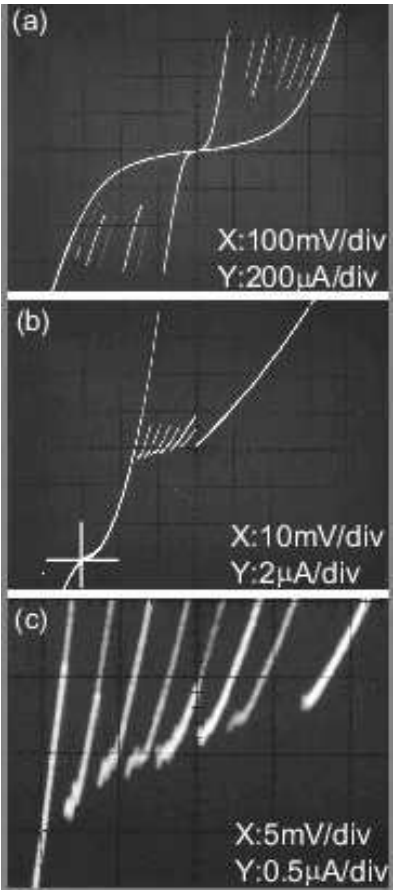}
\caption{The IVC of the sample Nm1-25 at 4.2 K. (a) Overall IVC. (b) Same IVC in the low bias region. (c) IVC on expanded scales showing the BPR.}\label{fig1}
\end{figure}

Figure \ref{fig1}(a) shows the typical IVC of the fabricated mesa (Nm1-25).
The IVC exhibits a multiple branch structure with large hysteresis.
From the number of branches, it is found that the mesa contains of 8 IJJs.
The critical current $I_c$ of constituent IJJs is nearly equal to each other and is 640~$\mu$A, corresponding to the current density of $J_c=2160$~A/cm$^2$.
The voltage spacing between branches at $I=I_c$ is $\sim30$mV.
Figure \ref{fig1}(b) shows the enlarged view of the IVC in the low bias region.
We note that the minimal currents of each branch, i.e., the value of current at jump point $I_j$, are not identical in contrast to the good homogeneity of $I_c$'s.
This feature is consistent with the prediction of the CCJJ+DC model.\cite{shukrinov:physicaC2006,shukrinov:sust2007,shukrinov:prb2007,shukrinov:prl2007}
The ununiformity of the $I_j$'s can also be caused by self-heating but the self-heating effect can be ignored for this sample because the voltage spacing between branches at $I_c$ is independent of the number of junctions in the voltage state in the stack.
Therefore, the observed ununiformity is predominantly due to the coupling between adjacent IJJs in the stack.
Furthermore, the branches demonstrate some structure before transition to another branch.
A detailed structure before $I_j$, which was obtained by sweeping the bias current up and down repeatedly, is shown in Fig. \ref{fig1}(c).
This is very similar to the BPR predicted in Ref.~7.
It also looks like the subgap structure caused by the resonance of the Josephson oscillations and phonon modes.\cite{schlenga:prl1996,seidel:physicaC1997,oya:physicaC2001}
However, we can rule out this phonon excitation because it should appear at $V\approx3,~6,~8$~mV and these voltages are inconsistent with our results.
Therefore, we consider that the observed structure is the experimental evidence of the BPR.

As it was shown in Ref. 8, in the CCJJ+DC model the hysteresis of IVC depends on the coupling parameter $\alpha=\varepsilon r_D/(d_sd_i)$ and the dissipation parameter $\beta=1/\sqrt{\beta_c}$, where $\varepsilon$ is the dielectric constant, $d_s$ is the thickness of superconducting layer, $d_t$ is the thickness of insulating layer and $\beta_c=2eCR^2I_c/\hbar$ is the McCumber parameter ($C$ : junction capacitance, $R$ : junction resistance, $\hbar$ : Plank's constant).
If $\alpha=0$ (no coupling) or it is negligible small, the IVC is consistent with that obtained in the resistively and capacitively shunted junction (RCSJ) model.
Therefore, we compare the hysteresis of the experimental IVCs with results of the RCSJ model.
However, for IJJs at low temperature the simple RCSJ model with constant $R$ might not be used because of strongly nonlinear IVC, as is seen in Fig. \ref{fig1}(a).
Thus, the IVC at 77 K, which becomes almost linear in a small voltage range,  was examined.

\begin{figure}
\includegraphics[width=0.29\textwidth]{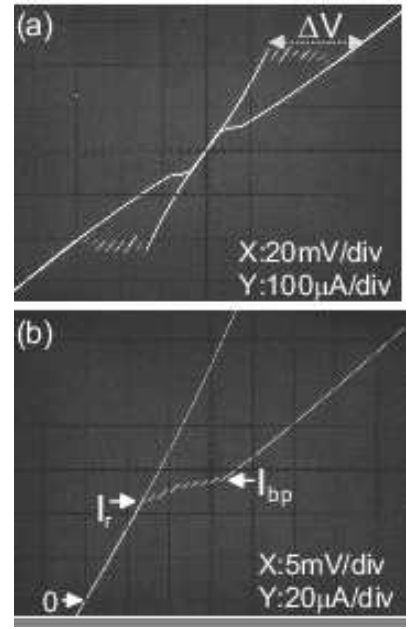}
\caption{The IVC of the sample Nm1-25 at 77 K. (a) Overall IVC. (b) IVC in the low bias region.}\label{fig2}
\end{figure}

Figure \ref{fig2}(a) shows the IVC of the mesa shown in Fig. \ref{fig1}, measured at 77 K and Fig. \ref{fig2}(b) shows its low bias current region.
As we can see, the multiple branches are almost linear but the large hysteresis still remains.
And it is found that even at 77 K $I_c$'s of the IJJs in the stack are almost identical.
In contrast, the $I_j$ of each branch is not uniform, as is seen in Fig. \ref{fig1}(c).
However, we couldn't observe the BPR at 77 K.
This is maybe because the thermal fluctuation makes the resonance a delicate state.
Therefore, in this case, $I_{bp}$ and $I_j$ coincide with each other.
In Fig. \ref{fig2}(b), we indicate the $I_{bp}$ on the outermost branch and the return current $I_r$ from the first branch.
Because the outermost branch extrapolates to zero current at zero voltage, from the sum voltage jump of $\Delta V=39.1$~mV at $I=I_c(=240~\mu$A) and $N=8$ one can estimate that the junction resistance $R$ of each IJJ is 20.4~$\Omega$ at 77~K.
Additionally, the junction capacitance $C=\varepsilon S/d_t$ is calculated to be 1.84 pF by assuming $\varepsilon_r=10$.
Substituting these parameters, we obtain $\beta_c(77~{\rm K})\approx560$ for the presented mesa.
Therefore, for the RCSJ model ($\alpha=0$) we expect $I_r^{RCSJ}\simeq13~\mu$A by using the relation $I_r/I_c=4/(\pi\beta_c^{-1/2})$ for $\beta_c>>1$.\cite{likharev}
This value is significantly different from the $I_r=45~\mu$A obtained from Fig. \ref{fig2}(b).
Such a difference between experimental $I_r$ and $I_r^{RCSJ}$ can be explained by the capacitive coupling  as shown in Ref. 8. 
Although another possible reason for the difference is the influence of thermal fluctuations,\cite{krasnov:prb2007} it seems unlikely that the fluctuations lead to such large difference discussed above.

\begin{figure}
\includegraphics[width=0.29\textwidth]{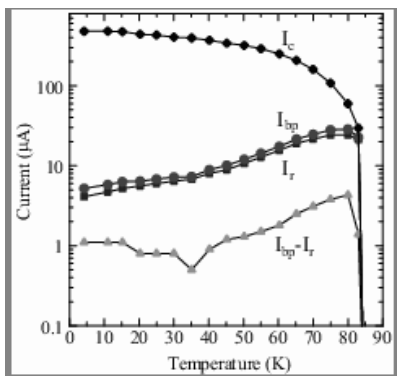}
\caption{Temperature dependence of $I_c$, $I_{bp}$, $I_r$ and $I_{bp}-I_r$ of the sample Nm5-25 ($S=25~\mu$m$^2$ and $N=8$).}\label{fig3}
\end{figure}

Figure \ref{fig3} shows the temperature dependence of $I_c$, $I_{bp}$, $I_r$ and $I_{bp}-I_r$ for a different mesa (Nm5-25) with $S=25~\mu$m$^2$ and $N=8$.
The $I_c(T)$ is slightly deviated from the Ambegaokar-Baratoff dependence due to the effect of thermal fluctuation on the switching current.
The $I_r$ hardly depends on $T$ at low temperature ($T<40$ K) and gradually increases with increasing temperature up to 80 K.
Such behavior is typical for IJJs.\cite{okanoue:apl2005,krasnov:prl2005}
It is also found that the $I_{bp}$ shows $T$ dependence similar to the $I_r$.
Furthermore, we note that $I_{bp}-I_{r}$ is almost independent of $T$ for $T<40$~K but at higher temperatures it increases with $T$.
As shown in Ref. 7, the $I_{bp}$ depends on $\alpha,~\beta$ and the wave number of excited LPW.
Assuming $\alpha$ is independent of $T$, the increase in $I_{bp}-I_r$ may arise from the change of the excited plasma mode.
However, to prove this, a further experiment would be required.

\begin{figure}
\includegraphics[width=0.29\textwidth]{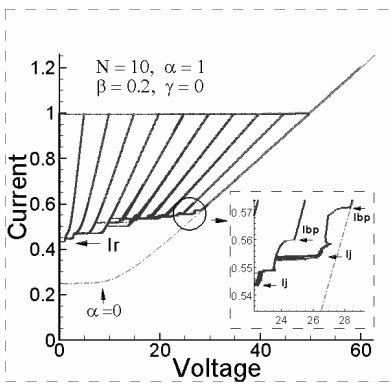}
\caption{The simulated IVC of a stack of 10 IJJs at $\alpha=1,~\beta=0.2$ and $\gamma=0$. The inset shows the enlarged BPR for the last two branches in IVC.}\label{fig4}
\end{figure}

In the framework of the CCJJ+DC model, the phase dynamics of IJJs can be described by the system of equations
\begin{eqnarray}
\frac{d^2\varphi_{\ell}}{dt^2}=I-\sin\varphi_{\ell}-\beta\frac{d\varphi_{\ell}}{dt}+\alpha(\sin\varphi_{\ell+1}+\sin\varphi_{\ell-1}\nonumber\\
-2\sin\varphi_{\ell})+\alpha\beta\left(\frac{d\varphi_{\ell+1}}{dt}+\frac{d\varphi_{\ell-1}}{dt}-2\frac{d\varphi_{\ell}}{dt}\right)
\end{eqnarray}
for the gauge-invariant phase difference $\varphi_{\ell}(t)=\theta_{\ell+1}(t)-\theta_{\ell}(t)-\frac{2e}{h}\int_{\ell}^{\ell+1}dzA_z(z,t)$ between superconducting layers in the stacks with $N$ IJJs.
Here $\theta_{\ell}$ is the phase of the order parameter in S-layer $\ell$, $A_z$ is the vector potential in the barrier.
Figure \ref{fig4} shows the simulation result of the IVC for a stack of 10 IJJs at coupling parameter $\alpha=1$, dissipation parameter $\beta=0.2$ and with nonperiodic boundary conditions $\gamma=0$, where $\gamma=s/s_0=s/s_N$ and $s$, $s_0$, $s_N$ are the thickness of the middle, first, and last superconducting layers, respectively.
In this figure, the dashed line corresponds to the IVC for $\alpha=0$.
It is found that Fig. \ref{fig4} replicates the experimental observations descried above, such as the equidistant branch structure, BPR, discrepancy of $I_j$ and the suppression of the hysteresis.

In summary, we have experimentally investigated the influence of the coupling between IJJs on the IVC of Bi$_2$Sr$_2$CaCu$_2$O$_y$.
We observed clear breakpoint region on the branches in the IVC and found that the coupling leads to the reduction of the hysteresis of the IVC at higher temperatures.
Our results are in a good agreement with recent theoretical work based on the CCJJ+DC model.


\end{document}